\documentclass{ws-mpla}

\begin{document}

\markboth{Donmez, O.}
{QPOs in numerical results and comparison with observations}

\title{Quasi Periodic Oscillations (QPOs) and frequencies in an accretion disk and comparison 
with the numerical results from non-rotating black hole computed by the GRH code}

\author{Orhan Donmez}

\address{Nigde University Faculty of Art and Science, Physics Department, \\
Nigde, Turkey 51200\footnote{electronic address:odonmez@nigde.edu.tr}} 

\maketitle

\pub{Received (Day Month Year)}{Revised (Day Month Year)}

\begin{abstract}
The shocked wave created on the accretion disk after different physical phenomena (accretion 
flows with pressure gradients, star-disk interaction etc.) may be responsible observed Quasi 
Periodic Oscillations (QPOs) in $X-$ray binaries. We present the set of characteristics 
frequencies associated with accretion disk around the rotating and non-rotating black holes 
for one particle case. These persistent frequencies are results of the rotating pattern in 
an accretion disk. We compare the frequency's from two different numerical results for fluid  flow 
around the non-rotating black hole with one 
particle case. The numerical results are taken from our papers Refs.\refcite{Donmez2} and \refcite{Donmez3} 
using fully general relativistic hydrodynamical code with non-selfgravitating disk. While the first 
numerical result has a relativistic tori around the black hole, the second one includes one-armed 
spiral shock wave produced from star-disk interaction. Some physical modes presented in the QPOs
can be excited in numerical simulation of relativistic tori and spiral waves on the accretion disk.
The results of these different dynamical structures on the accretion disk responsible for QPOs are 
discussed in detail.  

\end{abstract}

\ccode{PACS Nos.: include PACS Nos.}


\section{INTRODUCTION}
\label{INTRODUCTION}

Observed properties of the high frequency from $\approx 1$ Hz to $\approx 1$ kHz quasi periodic oscillations 
in the X-ray light curve of the black hole and the neutron star binaries have been studied by \cite{vander1}
and \cite{McCalum1} and references therein.  The studies of high frequency variability in $X-$ray binaries
provide a unique opportunity to explore the fundamental physics of space time and matter. The dynamical 
timescale on the order of several milliseconds is a timescale of the motion of matter through the region
located in close to black hole. Thus, Theories of the physics of the inner accretion disk, the orbital motion
of matter in relativistic gravity and the relativistic disk oscillations can be probed by high frequency signals 
from black hole binaries. The observation of high frequency properties have been motivated the theory of relativistic 
``diskoseismology'' which  studies the hydrodynamical oscillation modes of geometrically thin accretion disk 
for one particle case \cite{Wagoner1}, \cite{Kato1}.  Different types of shock waves, standing two-armed spiral
shock, moving one-armed shock and tori(\cite{Donmez2},\cite{Donmez3},\cite{Donmez5}, \cite{Donmez4}, \cite{Donmez1}), 
with discrete and different range of frequencies can exist in such disk
because the radial profile of general relativistic matters oscillation frequencies can create finite region where 
modes are trapped.

QPOs produced on the accretion disk in the black hole systems are thought to arise from physical processes. Depending 
on the where the oscillation reside, the various accretion models can be seen. These models are relevant to our 
understanding of various physical radii in the binary systems that are associated with high frequency variability 
according to the some theories. Some of these models accommodate movements of the inner accretion disk that are 
proposed to explained the  change in the frequency of high frequency QPOs. 

However, radial pressure gradients themselves can permit the existence of discrete modes because they produce 
a disk, or torus, of  finite extent. At least in the case of black hole $X-$ray binaries, QPO s are only 
observed in the power law spectrum states in which the flow is not solely composed of a geometrically thin
accretion disk. Therefore, it is believed that the radial pressure gradients may be responsible for trapping 
different frequencies modes on the accretion disk. Those reported have recently explored low 
frequency\cite{Giannios1} and high frequency\cite{Lee1},\cite{Kluzniak1} that accretion disk may 
explain observed QPOs.
Geometrically thick tori may form in stellar collapse, and modes of dense tori around black holes have 
also been suggested as a detectable source of gravitational waves\cite{Zanotti1}

The first part of the paper provides the theoretical background on the physics of $X-$ray binary system frequency
variability which is presented in accretion disks around the black hole for one particle system. This frequencies 
arise rotating patterns in the accretion disk and shock waves created on it and driven by gravity of black hole.
Second part of paper compares the numerical simulation results computed from fully general relativistic hydrodynamics 
code with one particle case. We also report the different modes on numerical modeled accretion disk around the
non-rotating black hole. 
 

\section{Characteristic Disk Frequencies}
\label{Caharacteristic Disk Frequencies}

\begin{figure}[pb]
\centerline{\psfig{file=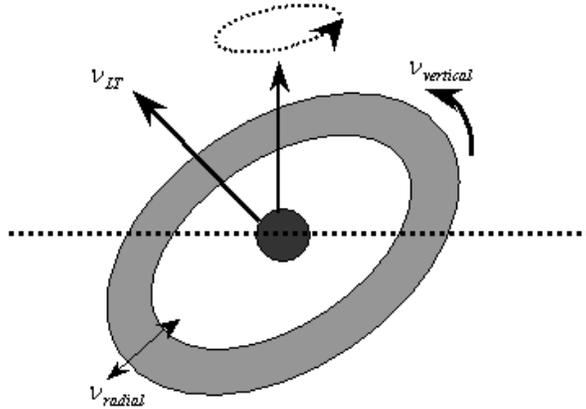,width=8cm}}
\vspace*{8pt}
\caption{Artistic visualization and representation of accretion disk black hole binary system
and their frequencies. }
\label{fig:1}
\end{figure}

The characteristic frequencies of this ring depend upon whether the spin axis of the ring
is tilted with respect to the spin axis of the black hole If we have a ring of matter orbiting 
a black hole that is itself spinning. Characteristic frequencies 
also depend upon the mass of the black hole, $M$, angular momentum of the black hole, $J$,
and the radius of ring, $r$. The artistic representation of the frequencies is shown in Fig.\ref{fig:1}.
The motion of a test particle in nearly circular orbits close to 
equatorial plane around a Kerr black hole can 
be decomposed into three components, which are circular planar motion at the orbital 
frequency $\nu_{\phi}$, harmonic radial motion at the radial epicyclic frequency $\nu_r$, and
harmonic vertical motion at the meridional epicyclic frequency $\nu_{\bot}$ \cite{Okazaki1}.  

A test particle  has an orbital frequency $\nu_{\phi}$ , measured by static observed at infinity\cite{Bardeen1}
called the Keplerian rotational frequency. This is the frequency at which matter rotates about the 
black hole, if the ring and black hole spin axis are aligned and given by

\begin{eqnarray}
\nu_{\phi} =  \frac{1}{2\pi} \left(\frac{M^{1/2}}{r^{3/2} +aM^{3/2}}\right).  
\label{eq:1}
\end{eqnarray}

\noindent
where the geometrized units $c = G = 1$ are used, $r$ is the orbital radius and $a$
is the dimensionless black hole spin parameter $(a = cJ/GM^2)$. We compute the theoretical and numerical 
calculation in dimensionless units scaled by black hole mass $M$ in the rest of the paper. 
The distances and times are measured in units of $GM/c^2$ and $GM/c^3$, respectively.

The radial epicyclic frequency $\nu_r$ is the frequency at which a particle oscillates about 
its abort if it is given radial 'kick'. The radial epicyclic frequency is zero at the 
inner edge of the disk, rises to a maximum near the disk inner edge, and then falls off
and is equal to Keplerian frequency at large radii. This falls off in frequency at the 
inner edge of the disk is a consequence of general relativity and strong field 
gravity. It can be define the depend on the observer far from disk  and given by

\begin{eqnarray}
\nu_{r}^2 = \nu_{\phi}^2 \left(1 - \frac{6M}{r} + \frac{8aM^{3/2}}{r^{3/2}} - 
\frac{3a^2M^2}{r^2}\right).
\label{eq:2}
\end{eqnarray}

The vertical epicyclic frequency $\nu_{\bot}$ corresponds the vertical perturbation of particle, rotating in an 
circular orbit around the black hole

\begin{eqnarray}
\nu_{\bot}^2 =  \nu_{\phi}^2 \left(1 - \frac{4aM^{3/2}}{r^{3/2}} + 
3\frac{a^2 M^2}{r^2}\right).
\label{eq:3}
\end{eqnarray}

\noindent
In Fig.\ref{fig:1}, $\nu_{LT}:$ the Lense-Thirring frequency is the frequency at which the spin 
axis of the ring and the spin axis of the black hole process about the total angular momentum, dominated
by the black hole, if these axis are not aligned. The Lense Thirring frequency is only 
comparable to the Keplerian orbital frequency for rapidly spinning black holes.

\begin{figure}[pb]
\centerline{\psfig{file=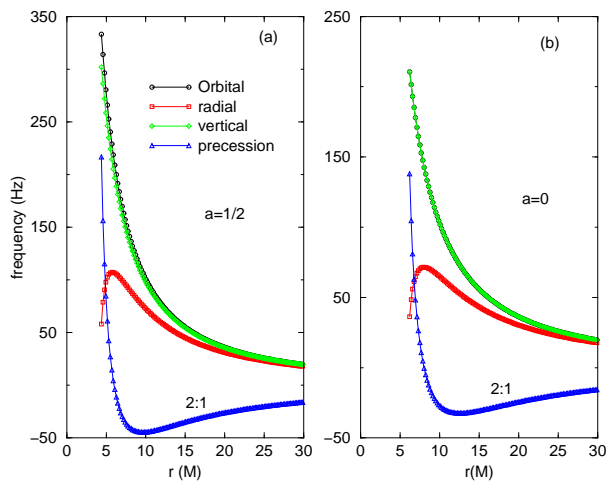,width=8cm}}
\vspace*{8pt}
\caption{Orbital frequency $\nu_{\phi}$, radial epicyclic frequency $\nu_r$, vertical
    oscillations frequency $\nu_{\bot}$ and
    precession frequency $\nu_p = \nu_{\phi} - 2\nu_r$ of a 2:1 orbit in an
    accretion disk around (a) a Kerr black hole  with $a=1/2$ and (b) a Schwarzschild black hole 
    for  $M = 10 M_{\odot}$.
    (a) Precession frequency exhibits a shallow negative minimum at
    $r_* \approx 9.65(142.1km)$ and the last stable circular orbit is located at 
    $r=4.21 (62.089km)$. (b) Precession frequency exhibits a shallow negative minimum at
    $r_* \approx 12.42(183.25km)$ and the last stable circular orbit is located at 
    $r=6.0 (88.48km)$. }
\label{fig:2}
\end{figure}

\begin{figure}[pb]
\centerline{\psfig{file=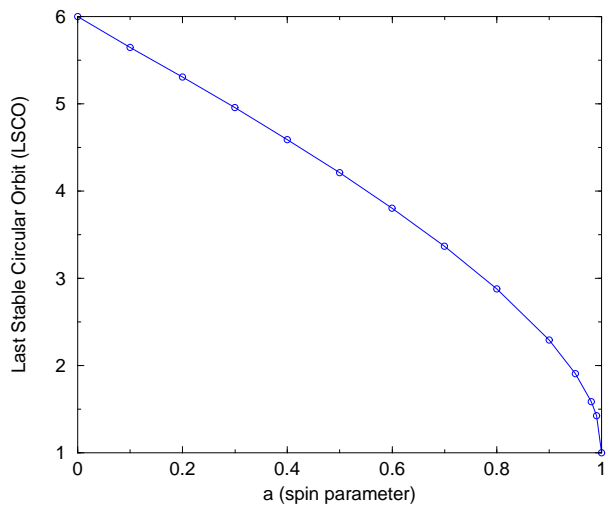,width=8cm}}
\vspace*{8pt}
\caption{The last stable circular orbit (LSCO) in the accretion disk around the 
Kerr black hole  vs. its spin parameter is plotted.}
\label{fig:3}
\end{figure}

In the Newtonian theory of spherically symmetric gravitating bodies (in the absent of Keplerian mechanics),
the three frequencies coincide ($\nu_{\phi} = \nu_{r} = \nu_{\bot}$) but in the relativistic theory 
$\nu_{\phi} > \nu_{r}$. The degeneracy of epicyclic frequencies are braked by general relativistic effects.
The all three frequencies are defined for rotating black hole. In case of non-rotating black hole, Schwarzschild, 
the frequencies are $\nu_{\phi} = \nu_{\bot} > \nu_{r}$. The radial dependences of these frequency for
a test particle in a rotating  and and non rotating black hole is illustrated in Fig.\ref{fig:2}. The 
vertical frequency equals to orbital one in case of non-rotating black hole. The rotating matter can be deviate 
around the minimum precession frequency, seen in Fig.\ref{fig:2}. 
Accounting a QPO frequency with a Keplerian frequency provides restriction on the relationship of 
the black hole mass  and the angular momentum for a given radius. The radius can be radius of marginally 
stable orbit, or any other physical radius of a region associated with the excitation of the QPO.  The circular 
orbits close to black hole are unstable and  last stable circular orbit (LSSO) is located where 
$\nu_{r}^2$ vanishes. As with the Schwarzschild black holes, the last stable orbits for a Kerr black hole 
are the circular orbits closest to the event horizon. Objects in orbit inside of the last stable orbit 
fall onto the event horizon. The Fig.\ref{fig:3} shows the LSSO as a function of spin parameter $a$ 
for clockwise orbit. LSSO tends to $1$ while the black hole spin goes to rapidly rotating black hole spin, $1$.

One of the fundamental  trapping oscillation has been identified by internal gravity, called the 
$g$ mode. The $g-$mode oscillations involve predominantly vertical displacements of the disk
and are mainly regulated by centrifugal and pressure gradient forces. 
The are trapped near the maximum of the radial epicyclic frequency in the resonant cavity,
seen in Fig.\ref{fig:2}. The $g$ mode should become trapped in the potential well of accretion
disk (geometrically thin) in a black hole potential. The depth and size of region depend where 
the modes are trapped depends on both the mass and the spin of the black hole.

\begin{figure}[pb]
\centerline{\psfig{file=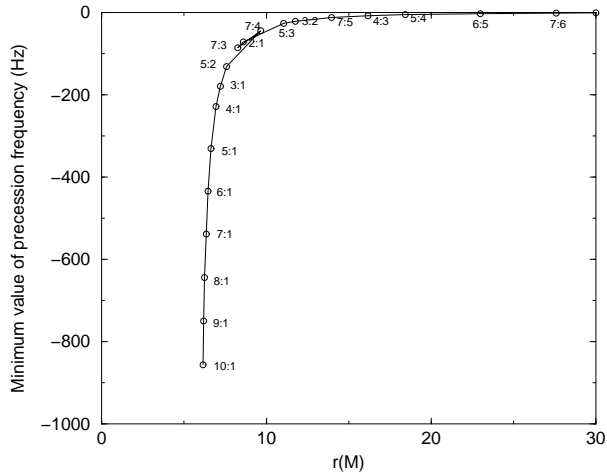,width=8cm}}
\vspace*{8pt}
\caption{Precession frequency exhibits a shallow negative minimum value for different orbits
in an accretion disk around a Kerr black hole  with $M = 10 M_{\odot}$ and $a=0.5$
While the rotation number increases in case of a fixed epicycle, the location of precession 
frequency decreases.} 
\label{fig:4}
\end{figure}

Black hole can make a close orbit in case of orbital and epicyclic 
frequencies are same. Because of general relativistic effects, the frequencies are in general different
and the orbits will precess. The orbits of particles that are eccentric and slightly tilted with respect 
to the radial are involved. On the other hand, particles in slightly tilted orbits fail to return to the 
initial displacement  from the equatorial plane in radial direction, after a full revolution around the
black hole. This introduces a precession frequency and define with $\nu_{prec} =  
\nu_{orb} - \zeta \nu_{rad}$ where $\zeta$ defines orbit, $n/m$ \cite{Mustafa1}. The integers $n$ and $m$ 
determine the shape of the precession orbit. Fig.\ref{fig:4}  plots the precession frequency as a 
function of radial coordinate for different mode. We have computed minimum values of precession 
frequency location.  The negative minimum value of precession gets bigger  and their location goes 
to black hole LSCO in case of bigger rotation number at fixed epicycle in an accretion disk around the 
Kerr black hole with the spin parameter $a=0.5$. In the relativistic precession models, QPO frequency variations
for a black hole are considered to be due to variations in the locations of the inner edge of the 
accretion disk.

Until now, we present the test particle epicyclic frequencies and their properties 
depend on black hole spin. We only focused on the kinematics and neglected particle interactions and 
the hydrodynamical effects of the disk. In a real accretion disk, the collective particle motion would 
have to be excited by some dynamical mechanism. It complicates and requires numerical simulations
of the disk to fully understand the driving process, and the effects of the hydrodynamical forces and pressure
jumps. In the next section, we present and discuss the oscillating properties of relativistic,
non-selfgravitating tori and one-armed spiral shock wave created in the accretion disk around the
black hole \cite{Donmez2},\cite{Donmez3}.


\section{Numerical Results}
\label{Numerical results}

The characteristic frequencies defined for one-particle system are also important in 
a discussion of fluid motion about a gravitating body in the general relativity\cite{Kato2}. 
The accretion disk is a body of hot gas in some regions that supported against infall 
primarily by rotation and is nearly hydrostatics equilibrium. The theories of accretion 
admit solutions in which the disk thickness is much smaller than its radial extent, but 
also solutions in which the disk has the geometry of a torus. Like other extended bodies 
in equilibrium, the disk capable of motion in a variety of modes. The radial vibrations 
of two-dimensional models of geometrically thick disk have been investigated numerically 
in the Schwarzschild black hole\cite{Rezzolla1}. 

It is pointed out that in general relativity, in addition to the  fixed frequencies, orbital, 
radial and vertical, there are other preferred frequencies, those of 2:1, 3:1, etc., 
resonances between orbital and radial epicyclic frequencies. These are possible because 
the ratio of orbital and radial epicyclic frequencies tends to large values near the 
marginally stable orbit: $\nu_{\phi}/\nu_{r} \to \infty$ as $r \to 6M$ for non-rotating 
black hole, seen in Fig.\ref{fig:2}. Frequencies in 2:1 or 3:1 ratio can be in 
resonance because epicyclic motion is enharmonic. As a usual for non-linear oscillators, 
the resonance occurs for a range of frequencies near the eigenfrequency of the oscillator, 
so the the orbital frequency need not be an exact multiple of the eigenfrequency of the 
epicyclic oscillator, nor need it be constant. Here, we show the different frequencies and 
their ratios produced from two different numerical results, toris created around the black hole
and one-armed spiral shock wave formed as a consequence of star-disk interaction.


\subsection{Torus created close to black hole}
\label{Epicylic Frequencies from Numerical Modelling}

To briefly summarize the basic properties of the oscillating torus model, 
we recall that we are considering a  non-selfgravitating perfect fluid toris orbiting around
the non-rotating black hole\cite{Donmez2}. The fluid is rotating in a circular non-geodesic motion 
and the equations of relativistic hydrodynamics in a Schwarzschild black hole space-time are solved 
using the general relativistic hydrodynamics code described in Ref.\refcite{Donmez6}. The introduction
of the harmonic oscillations of the torus having centrifugal and pressure-gradients as the restoring forces.
The restoring forces can explain toris
dynamic of accretion disk. A first restoring force is the gravitational force in the direction 
vertical to orbital plane and which will make  a harmonic oscillation on equatorial plane. Second 
force is provided by the pressure gradients. Third restoring force known as centrifugal force is 
responsible for internal oscillations of the orbital motion of the disk and the marginally 
bound motion produces epicyclic frequencies.

\begin{figure}[pb]
\centerline{\psfig{file=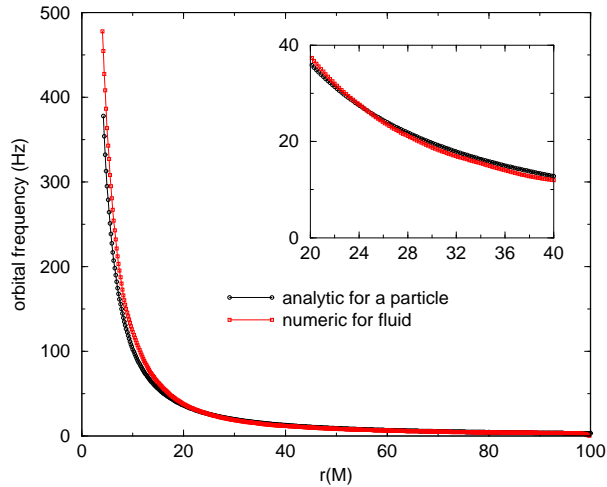,width=8cm}}
\vspace*{8pt}
\caption{Orbital frequencies plotted for analytic and numerical considerations as a function 
of radial dependence. Comparison of the orbital epicyclic frequency of  a test particle with
a perfect fluid disk case which is produced from fully general relativistic hydrodynamic code.
The numerical results used here are taken from Donmez, 2006 (Submitted), seen in Fig.3 at $t=43712.61M$ 
which shows the created accretion disk and tori around the non-rotating black hole with mass $M = 10 M_{\odot}$.} 
\label{fig:NA:1}
\end{figure}

\begin{figure}[pb]
\centerline{\psfig{file=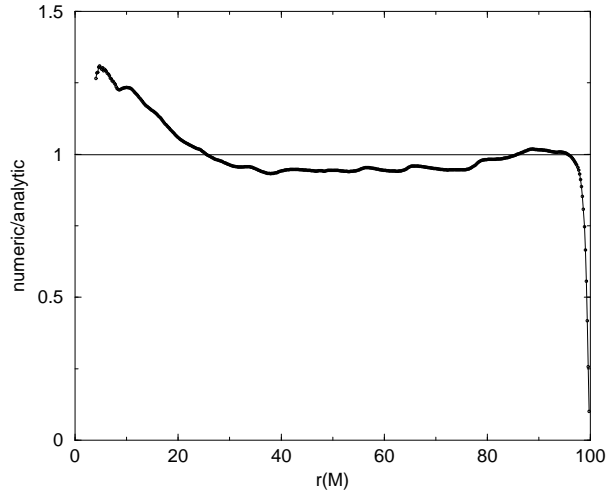,width=8cm}}
\vspace*{8pt}
\caption{Frequencies ratio of numerical result with analytic value as a function of $r$ 
in Fig.\ref{fig:NA:1}. If the ratio is bigger than $1$, it represents the super-Keplerian 
disk which is rotating faster than  Keplerian disk. Otherwise disk is sub-Keplerian. 
$r = 25.37M$ represents the location where Keplerian velocity equal to orbital 
velocity of the fluid. Inside that radius, the disk has super-Keplerian angular velocity. The  
sub-Keplerian disk becomes puff-tuffed and geometrically thick in place where 
the disk is super-Keplerian. } 
\label{fig:NA:2}
\end{figure}

\begin{figure}[pb]
\centerline{\psfig{file=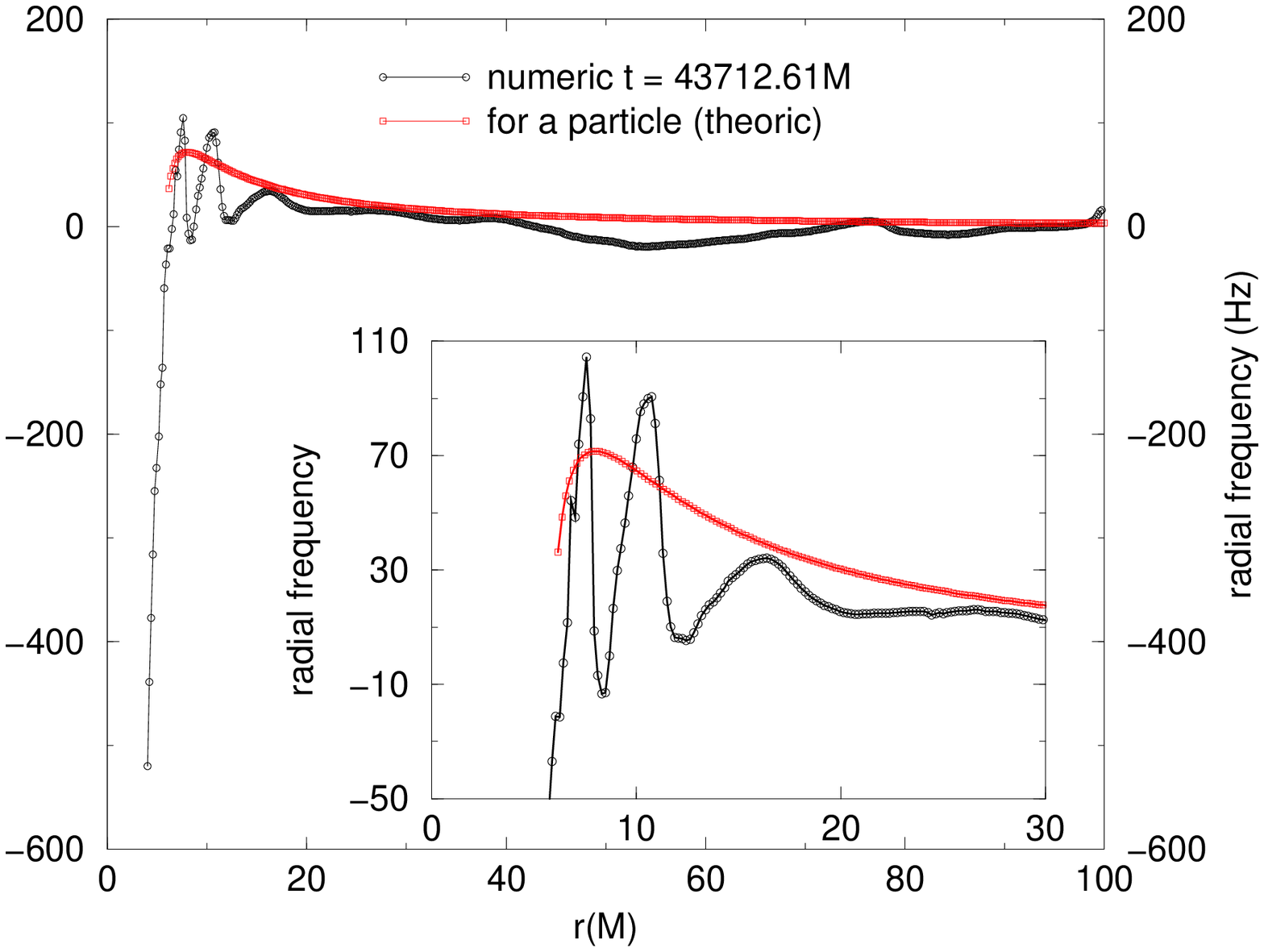,width=8cm}}
\vspace*{8pt}
\caption{Radial frequencies of the theoretical and the numerical results (same as  
Figs.\ref{fig:NA:1} and \ref{fig:NA:2}) as a function of radial coordinate. The numerical 
and theoretical results can be compared. The numerical radial frequency is oscillating 
close to black hole. The last stable circular orbit for both cases almost in same place. 
The fluid falls into black hole and  goes faster and faster and nothing can escape to the 
outside world, not even light when the matter closes to black hole (horizon, $2M$) in 
numerical simulation. On the other hand, particle radial frequency goes zero 
when particle reaches to last stable circular orbit for non-rotating black hole, 
$6M$. The radial epicyclic frequency  reaches its 
maximum  at $r=8M$ in theoretical consideration and $r=7.6M$ in numerical 
calculation for a non-rotating black hole and both go to zero at last stable 
circular orbit. This figure shows a luminous torus oscillating along its own axis.} 
\label{fig:NA:3}
\end{figure}

In a real accretion disk, the collective particle motion would have to
be excited by some dynamical mechanism. It complicates and
requires numerical simulations of the disk to fully understand the
driving process. In Fig.\ref{fig:NA:1}, we give the results of vertical epicyclic frequency
from numerical simulation of accretion disk around the Schwarzschild black hole using
General Relativistic Hydrodynamic (GRH) code\cite{Donmez2}, \cite{Donmez1} 
and a test particle. In the test particle case,
we content with studying the kinematics only. We set off a collective mode by selecting
appropriate initial conditions  and follow the pattern evolution by tracing the motion 
of one particle. Self interaction of particle and hydrodynamical effects are neglected.
In the numerical simulation case, we have all effects except the self gravity.  
Contrast to Ref.\refcite{Lei1} for neutron star, it is 
seen in the Fig.\ref{fig:NA:1} that orbital epicyclic frequency in a perfect fluid disk
case is sup- or super-Keplerian depends on the radial location for the same black hole mass, 
seen in Fig.\ref{fig:NA:2}. 
So the deviation of the these frequencies using same black hole may come from the 
omission of hydrodynamic effect for a test particle. 
Fig.\ref{fig:NA:1} shows the orbital frequency of toris rotating around the Schwarzschild black hole of 
mass $M = 10 M_{\odot}$ as a function of radial coordinate. We find that the torus performs harmonic oscillations
both in the radial and vertical directions. Both these oscillations speak of a non-linearity in the system.
Orbital frequency for a test particle and numerical
result for fluid are plotted on the same graphics and they are comparable. 
The fluid in the accretion disk has rotated oscillation frequency 
depends on the radial dependence. While the matter inside the $r=28M$ rotates with super-Keplerian frequency, it is 
sup-Keplerian in the outside this radius, seen in Fig.\ref{fig:NA:2}. The disk which rotates super-Keplerian frequency 
has strong centrifugal force and it is balanced by strong gravity close to black hole. As a consequence of this phenomenon,
the tori is crated around the black hole. The gas pressure dominates radiation
pressure throughout the disk except (depending upon parameters) in the innermost regions where the 
temperatures are high  and it is outside the $r=28M$.  Fig.\ref{fig:NA:3} shows the radial frequency from 
numerical result of tori and a test particle case. They have a similar behavior but numerical results have
more oscillatory motions because of some forces which would not produced in the test particle case and numerical errors.
The numerical errors are produced using with low resolution in numerical simulations. Using the high resolution is not
possible in this case because we have to wait until $t \approx 37000M$ to reach steady state.
The strong gravity is responsible for the presence of two frequencies $\nu_{\phi} \neq \nu_r$,
seen in Figs.\ref{fig:NA:1} and \ref{fig:NA:3}, where in Newtonian gravity there is only one 
frequency ($\nu_r = \nu_{\phi} =  \nu_{\bot}$). 

The transition from a Keplerian to a sub-Keplerian flow may proceed smoothly, although 
it is very likely that a perfect adjustment never occurs. In general, the transition 
should take place through the setting up of a centrifugal barrier  (where a centrifugal 
force slightly exceeds the gravitational force) within the adjustment radius. Inside potential 
barrier, matter can be hold and produces a tori close to black hole. 
Here we discuss the formation of kinks and shocks in the supersonic regime of accretion flow (in tori) 
as a possible physical reason for a super-Keplerian rotation. In a 
region, seen in Fig.\ref{fig:NA:2}, with super-Keplerian rotation matter may experience the relaxation oscillations 
in radial directions. These oscillations are expected to be in a resonance with the 
local angular velocity in a disk, and the variation in an emitting area caused by 
the oscillations around a transition point produces the quasi-periodic oscillations 
(QPOs) in the X-ray flux. The relatively soft disk photons in a centrifugal barrier
region are scattered off the hot electrons thus forming the Comptonized X-ray spectrum.
The electron temperature is regulated by the supply of soft photons from a disk, 
which depends on the ratio of the energy release (accretion rate) in a disk and 
the energy release in centrifugal barrier region. For example, the electron 
temperature is higher for lower accretion rates while for a high accretion rate 
(of order of the Eddington one) a centrifugal barrier region cools down very 
efficiently due to Comptonization.

\begin{figure}[pb]
\centerline{\psfig{file=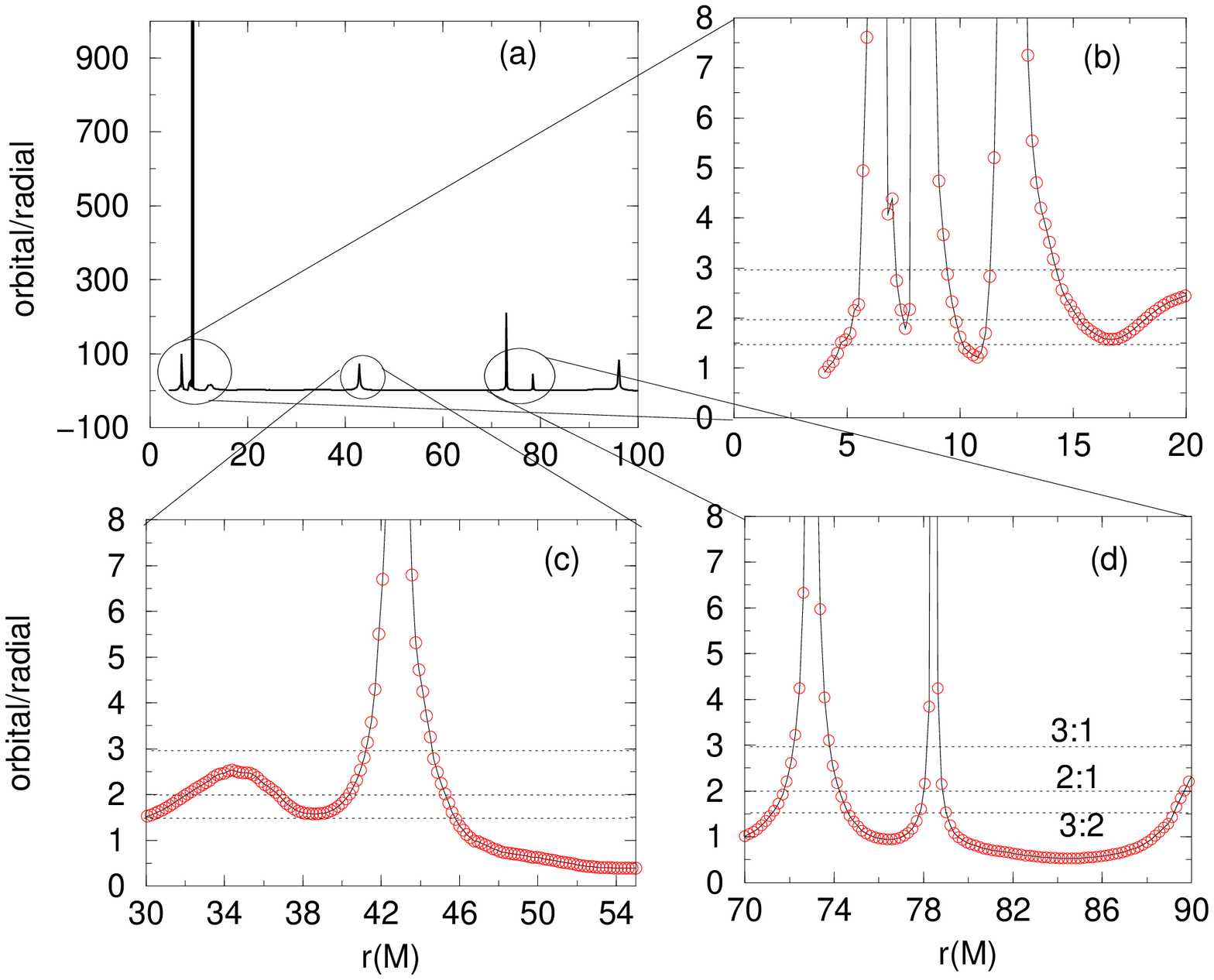,width=8cm}}
\vspace*{8pt}
\caption{The ratio of orbital to radial frequency to find out particular radius that corresponds to
$2:1$, $3:1$ and $3:2$  orbits, which are representative of the deformed and the self-intersecting orbit 
classes, in numerical calculation for a perfect fluid. The perfect fluid disk case produced from 
fully general relativistic hydrodynamic code. The numerical results used here are taken from 
Donmez, 2006 (Submitted), seen in Fig.3 at $t=43712.61M$ which shows the created accretion disk 
and tori around the non-rotating black hole. The mass of the black hole is $M = 10 M_{\odot}$.
(a) Represents the ratio of orbital to radial frequency as a function of radial coordinate. 
(b) Represents the same ration but focused around $0 < r < 20$. $3:2$ orbits found  at 
$r \approx 4.6M$, $r \approx 10.11M$, $r \approx 11M$ and $r \approx 16.67M$. $2:1$ orbits 
found  at $r \approx 5.1M$, $r \approx 7.5M$, $r \approx 9.7M$, $r \approx 11M$, $r \approx 15.2M$,
and $r \approx 18.18M$. $3:1$ orbits found  at $r \approx 5.45M$, $r \approx 6.9M$, 
 $r \approx 7.74M$, $r \approx 9.3M$, $r \approx 11.22M$, and $r \approx 14M$. The orbits 
radial locations in the (c) and (d) can also be seen. \label{fig:Num:1}}
\end{figure}

The axisymetric oscillation modes of relativistic tori consists of different modes.
Parametric resonance between two oscillations is possible in accretion disk because 
the coupling of modes is non-linear in hydrodynamics.
The ratio of orbital velocity to radial frequency is given in Fig.\ref{fig:Num:1}.
The radial mode oscillation as the origin of high frequency QPOs in black hole systems has been
observed in the different frequency ratios $1:2:3....$ at different radial locations at a fixed $t=43712.61M$. 
The frequency ratios $5:3$ and $3:2$ have also been noted in Refs.\refcite{Abrom1} and \refcite{Remillard1}.

\begin{figure}[pb]
\centerline{\psfig{file=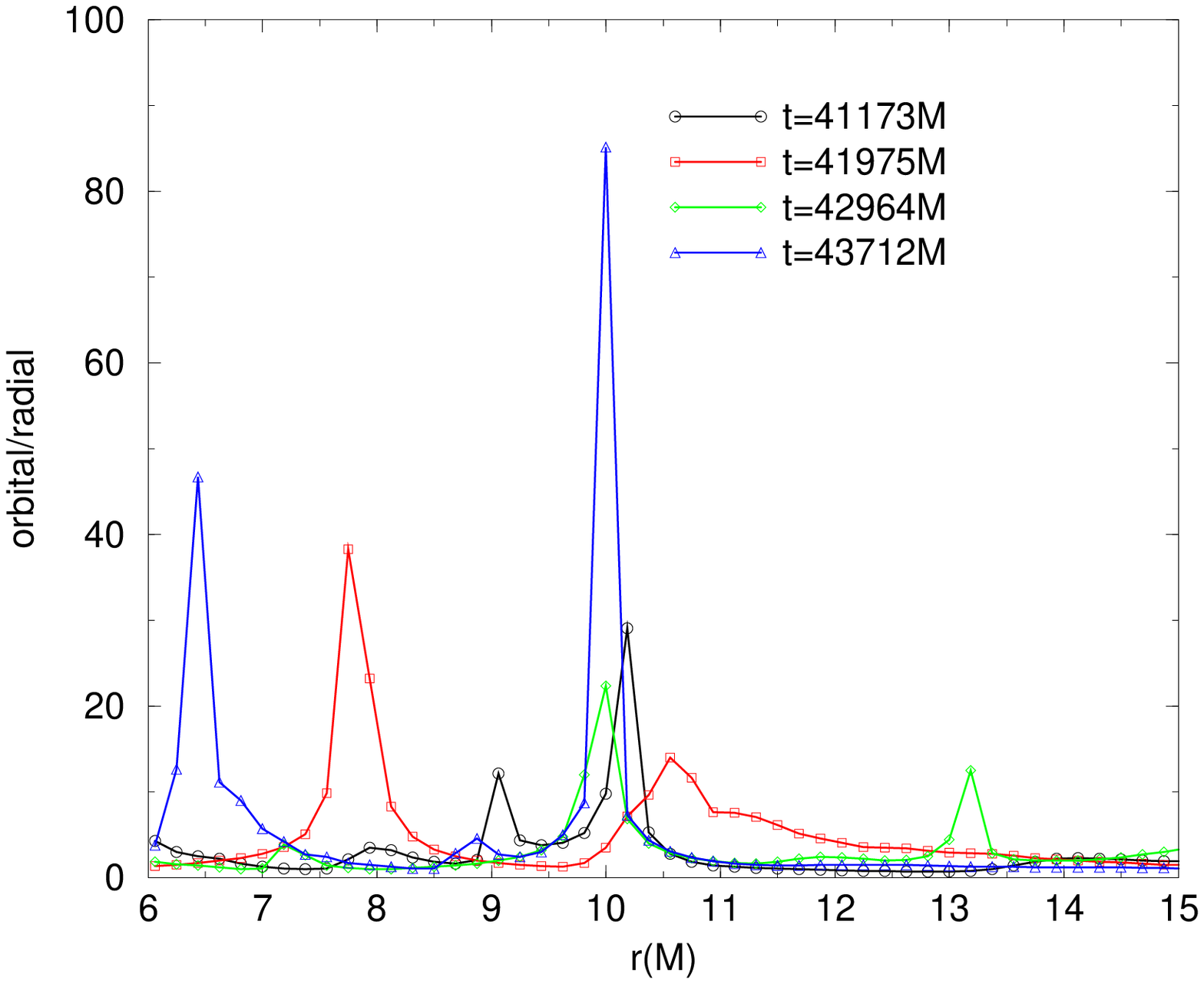,width=8cm}}
\vspace*{8pt}
\caption{Shows the time evolution of ratio of frequencies computed from numerically modeled accretion disk
which included tori around the black hole at 
a fixed radial distance,  The location of different  ratios
stay in one place with the small oscillation around the $10.2M$ There are some other rations which can also be seen 
but they are not stay at a fixed $r$. The variation of ration at  $10.2M$ in different times shows the vertical
oscillation of torus around the non-rotating black hole. The lower of the frequencies may reflect changes in  the 
emissivity  of the torus.} 
\label{fig:NA:3:1}
\end{figure}

The investigations of the oscillatory properties of non-selfgravitating  tori orbiting around the black hole
and their frequency ratios for different modes at different times are shown in Fig.\ref{fig:NA:3:1}
after tori reached to steady state. 
The location of different ratios stay in one place with small oscillation around the $10.2M$.  It produces 
consistent high frequency and it can be observed different $X-ray$ detectors. Another important thing it 
can be seen here that
twin high frequency rations produced at each time step but they produce oscillatory behavior, 
seen in Fig.\ref{fig:NA:3:1}. For the three microquasars GRO J1655-40, XTE J1550-564 and GRS 1915+105 twin 
high frequency quasi-periodic oscillations (HFQPOs) with a ratio of $3:2$ and $3:1$ 
have been measured\cite{Aschenbach1}.

\subsection{One-armed spiral shock wave created in case of star-disk interaction}

The perturbed accretion disk around the non-rotating black hole created rotating one-armed spiral 
shock wave. This spiral shock wave transports the angular momentum out of disk  when the matter 
collides with it and it causes the gas falling into the black hole\cite{Donmez5}. Dynamically stable shock
wave generally  oscillates in a mixture of internal and global modes. Internal modes cause oscillations of
the pressure and density profile at the shock front. The reflected flux is therefore directly modulated
by the changes in the thermodynamical properties of the gas, while the dynamic of rotating shock wave is 
nearly unchanged. 

\begin{figure}[pb]
\centerline{\psfig{file=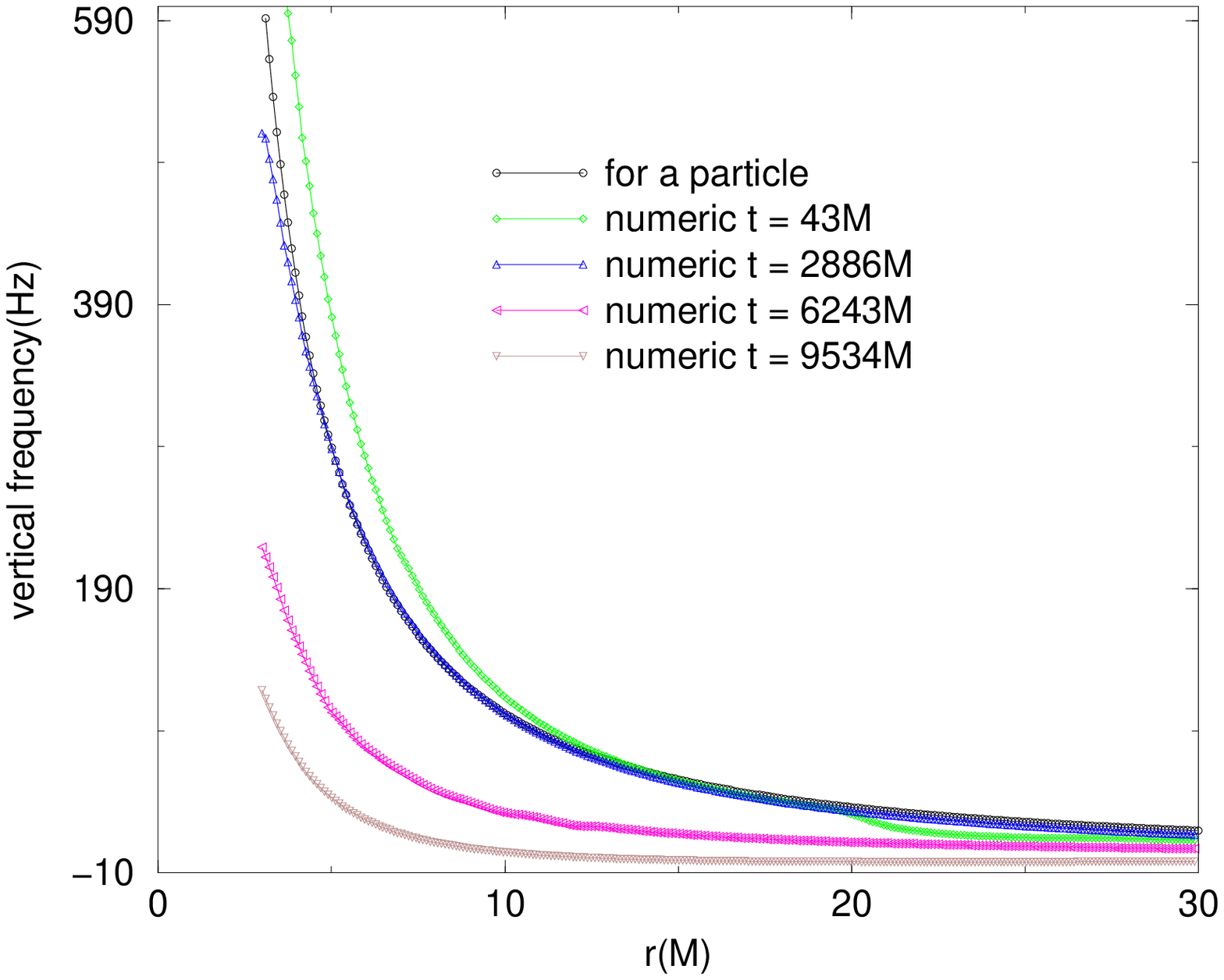,width=8cm}}
\vspace*{8pt}
\caption{Orbital frequencies plotted for analytic and numerical considerations at different times 
as a function of radial dependence. 
Comparison of the orbital epicyclic frequencies of a test particle with
a perfect fluid disk case for star disk interaction processes produced from fully general 
relativistic hydrodynamic code. There is a one-armed spiral shock wave is created on the accretion disk,
seen in Donmez, 2006 (in press). It causes to loose to mass of black hole and reduces to epicyclic frequency of
accretion disk during the time evolution. For both cases, the mass of the black hole is 
$M = 10 M_{\odot}$ and $a=0.0$ which is the Schwarzschild black hole.} 
\label{fig:NA:4}
\end{figure}

The radial dependences of orbital (vertical) frequencies at different times and fixed angular distance, 
$\phi=0$ for a non-rotating black hole 
and accretion disk with rotating shock are  illustrated in Fig.\ref{fig:NA:4}. in the early time of 
simulations, orbital frequency of disk shows oscillatory behavior around the test particle 
orbital frequency and it diverges from the test particle case during the process when the gas is falling 
into black hole.  The gas on the disk rotates smaller oscillatory velocity and it produces smaller 
epicyclic frequency.

\begin{figure}[pb]
\centerline{\psfig{file=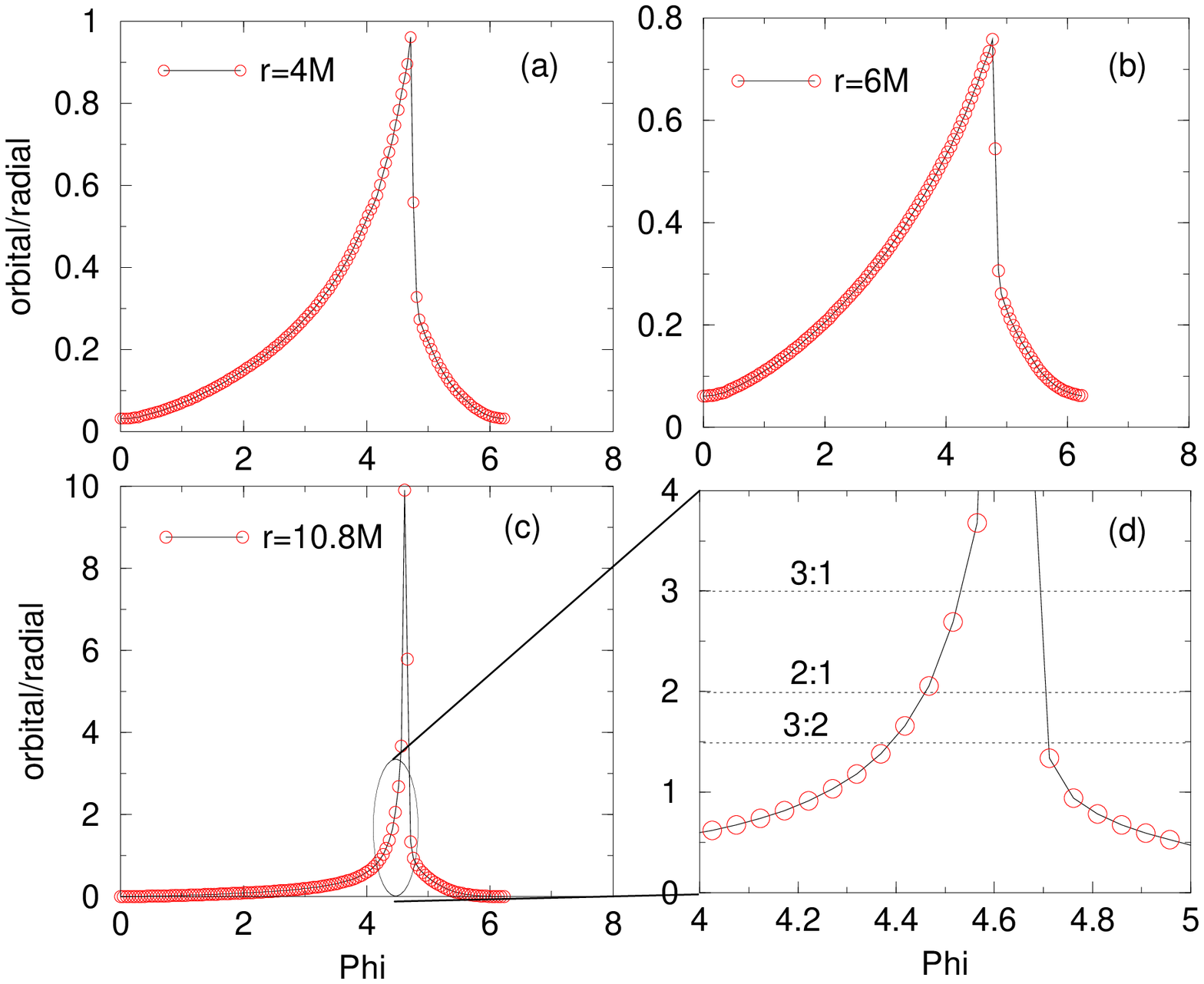,width=8cm}}
\vspace*{8pt}
\caption{The ratio of orbital to radial frequencies in the accretion disk which is produced 
by general relativistic hydrodynamics code as a results of star-disk interaction.
The numerical results used here are taken from Donmez, 2006 (in press), seen in Fig.3 at $t=9534.70M1M$
which shows the created one armed-spiral shock wave around the non-rotating black hole. The 
ration of frequency are plotted as a function of angular distance at a fixed $r$.
(a)shows the ratio at $r=4M$. It is inside the last stable circular orbit and the ratio is $< 1$.
(b)shows the ratio at $r=6M$.  It is on  the last stable circular orbit and the ratio is also $< 1$.
(c)shows the ratio at $r=10.8M$. It is outside  the last stable circular orbit and  the ratio has
$> 1$ values. It means that we have different orbits. (d) we have focused around $\phi=4.5$ where
there is a pick. We have also pointed the orbits $2:1$, $3:1$ and $3:2$.} 
\label{fig:Num:2}
\end{figure}

\begin{figure}[pb]
\centerline{\psfig{file=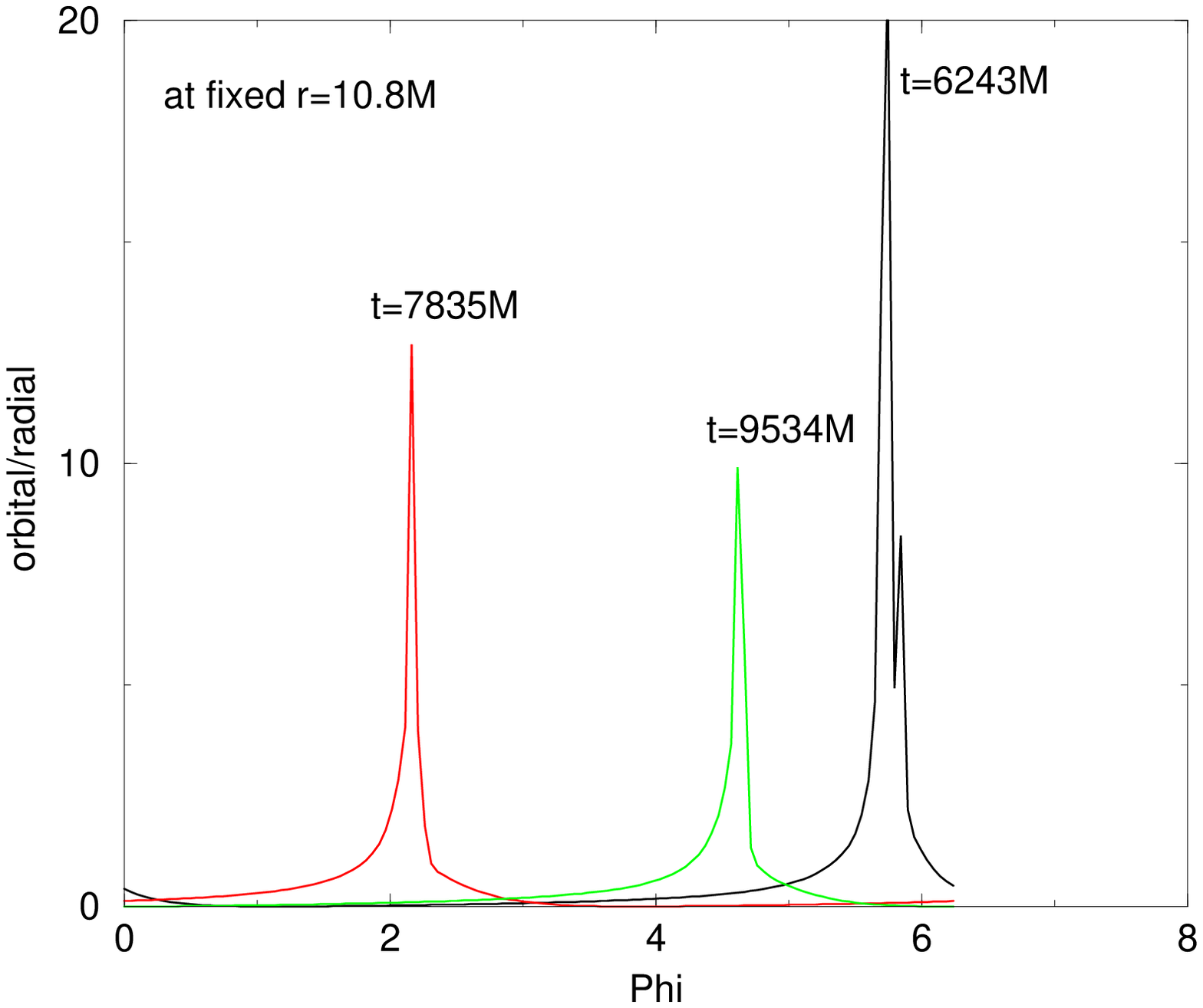,width=8cm}}
\vspace*{8pt}
\caption{Shows the time evolution of ratio of frequencies computed from star-disk interaction at 
a fixed radial distance, $r=10.8M$. The location of 3:2 ratios
stay in one place but it is moving in angular direction after one-armed shock wave created which
moves around the black hole. The times $6243M = 0.306$ second, $7835M=  0.385$ second and 
$9534M=  0.468$ second  calculated for the black hole, $M = 10 M_{\odot}$.} 
\label{fig:Num:3:1}
\end{figure}

Fig.\ref{fig:Num:2} shows the ration of orbital to radial frequency for one-armed spiral shock case 
at different radial fixed location at a snapshot, $t=9534.70M1M$. While the frequency ratio at $r=4M$ and 
$r=6M$ do not exhibit any commensurate frequencies, the frequency ratio at  $r=10.8M$ produces 
commensurate frequencies in different ratios, $3:1$, $3:2$, $2:1$ etc. In the particular radius, different 
ratios can be found in orbital and radial coordinate frequencies.
Commensurate QPO frequencies can be seen 
as a signature of an oscillation driven by some type of resonance conditions. Fig.\ref{fig:Num:3:1} shows the
location of different ratios at a fixed radial coordinate $r=10.8M$ at different times. The oscillatory behavior
of accretion disk produces consistent frequency ratio due to shock wave created. The location of 3:2 ratios
stay in one place but it is moving in angular direction after one-armed shock wave created which
moves around the black hole.

\begin{figure}[pb]
\centerline{\psfig{file=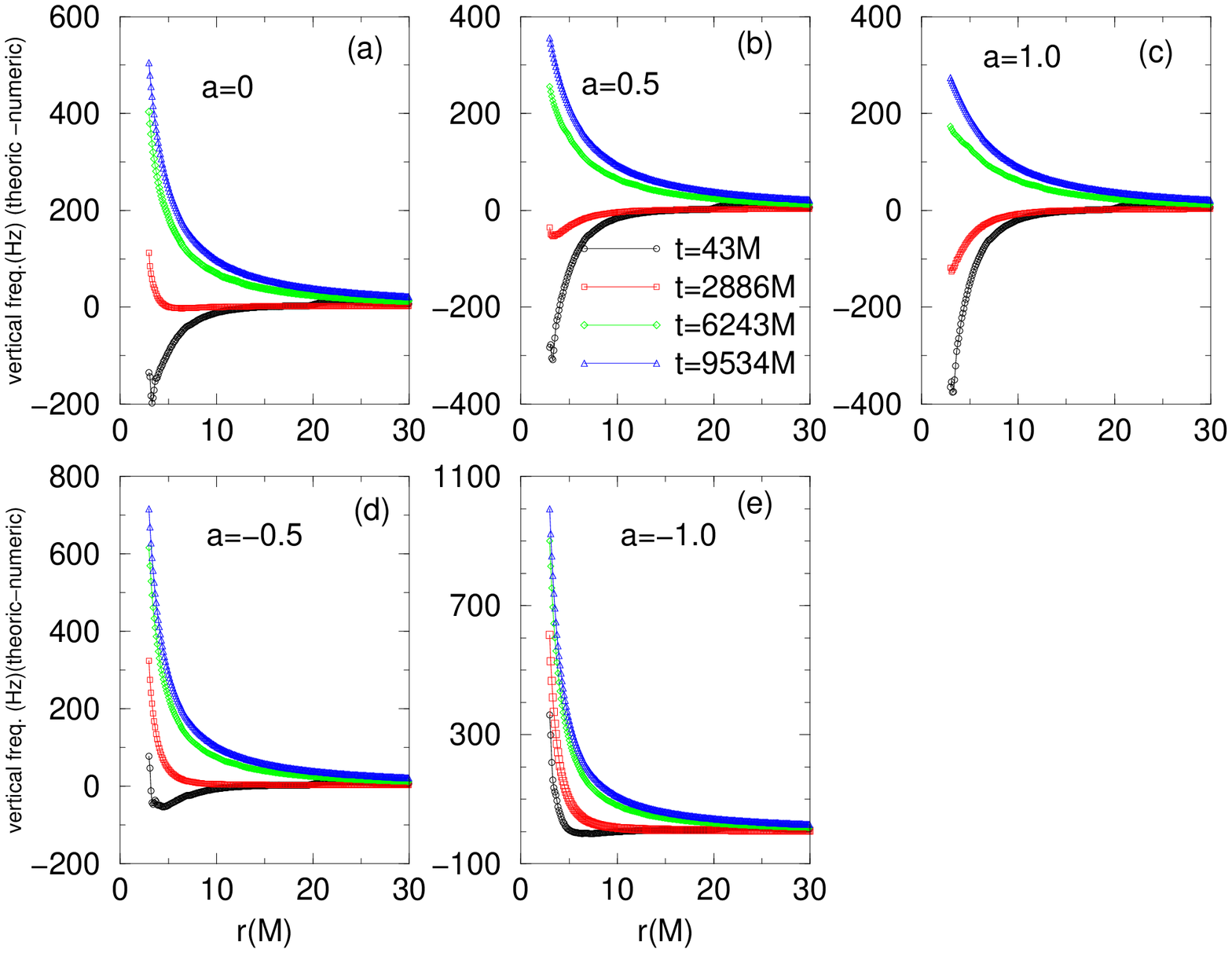,width=8cm}}
\vspace*{8pt}
\caption{Same data used in Fig.\ref{fig:NA:4} but difference between theoretical and numerical 
results are plotted as a function of $r$ at different times. } 
\label{fig:NA:5}
\end{figure}

Difference between theoretical results for a fixed spin parameter with numerical results for a non-rotating
black hole at different time steps are computed in Fig.\ref{fig:NA:5}.  The black hole and particles are corotaing 
orbits, where the black hole spin parameter is positive. On the other hand, the black hole 
and particle are the counterrotating orbits, while the black hole spin parameter is negative.
The numerical results are taken from in our paper Ref.\refcite{Donmez5} for the non-rotating 
black hole. Fig.\ref{fig:NA:5}(b)
and (c) show that results of spinning black hole vertical frequency tends to numerical values.
Because of tendency of numerical value to theoretical value for rotating black hole, 
even though we have stared with accretion disk problem around the 
non-rotating black hole, the black hole may have spinning because of the hydrodynamical effects.
But the gap between theoretical  and numerical results for the vertical epicyclic frequency in the the 
black hole which rotates in the counterrotating orbits gets bigger. This graphic gives big clue to
explain that the non-spinning black hole may rotate due to effects which are created interactions. 
The transfer of mass and angular momentum  from the disk to the black hole  
and interactions among them may lead to the gradual increase of the rotation law of the 
initially non-rotating  black hole, Schwarzschild.

\section{Comparison with Observations}
\label{Comparation with Observations}

Until now, we have given numerical and theoretical results of epicyclic frequency in black 
hole binary system. The numerical results have produced two-different models, tori created around the 
black hole and one-armed shock wave created on the accretion disk as a result of star disk interaction.
After this point, we compare the results from numerical calculations with observed one.
We make comparison for the presence in the black hole X-ray source of observed stable 
frequencies, demonstrated as narrow-with QPOs. These oscillation is produced as a result of parametric
resonance  in the accretion disk, and the observed frequencies ratios correspond to that resonance. 

In this section we discuss the comparison of high frequency QPOs detected by RXTE in black hole 
binary candidate with numerical founded frequencies. High frequency QPOs at $40-450$ Hz have been 
observed with RXTE for four black hole candidates. Furthermore, three of these sources show harmonic 
(3:2) pairs of frequencies. The models for these QPOs, orbital resonance and diskoseismic oscillations,
invoke strong gravitational effects. In the inner accretion disk and they depend on the black hole mass.
The vertical and radial frequencies ratios are given for numerical calculations in
 Figs.\ref{fig:NA:3:1} and \ref{fig:Num:3:1}. These figures represent different ratios, $3:2$,
$2:1$, $3:1$.... Some of the ratios are agreement with observed one.  There are, possibly, two 
distinct groups among  black hole binary candidate with the high frequency QPOs 
(in the $\sim 100-300$ Hz range). In GRS $1915+105$ and GRO $J1655-40$ the QPO frequency is stable.
The frequencies of the  GRO $J1655-40$ is the $450$ Hz for vertical and $300$ Hz for 
radial is has been argued by Ref.\refcite{Strohmayer1} that high frequencies
are sufficiently high that they require substantial black hole spin. For example, in the case of 
GRO $1655-40$ the $450$ Hz frequency exceeds the maximum orbital frequency at the innermost stable
circular orbit around the non-rotating black hole, seen in Fig.\ref{fig:NA:1}.
 The frequencies of the GRS $1915+105$ has recently been found in the 
relevant $3:2$ pair at $168$ for vertical and $113$ Hz for radial\cite{McClintock1}. 
Figures \ref{fig:NA:3:1} and \ref{fig:Num:3:1} computed from numerical calculations show the 
stable high frequency QPO in different ratios and high frequency signature can also  be a specific 
signature of a microquasar (both GRO $J1655-40$  and  GRS $1915+105$).

In two discovered transients, XTE $J1550-564$ and XTE $J1859+226$, the QPO frequency substantially 
changes from observation to observation. The frequencies of  XTE $J1550-564$ have been measure $276$ Hz 
for vertical and $184$ Hz for radial\cite{Remillard1}. Depend on the mass of observed black hole,
vertical and radial frequencies can be changed in XTE $J1859+226$ black hole binaries. 
Ref.\refcite{Cui1} report the measurement of a single high frequency QPO at $187$ Hz associated with 
XTE $J1859+226$. If the $187$ Hz frequency corresponds to the vertical epicyclic $3:2$ resonance, 
the second high frequency QPO at $125$ Hz, which is the radial epicyclic of oscillation. On the other side,
if the $187$ Hz frequency  is the result of radial epicyclic frequency, the expected vertical epicyclic 
frequency is at $281$ Hz.  These sources also exhibit pairs of QPOs that have 
commensurate frequencies in a $3:2$ ratio. 

Commensurate high frequency QPO frequencies can be seen as a signature of  an oscillation driven by some
type of resonance conditions. Ref.\refcite{Abramowicz1} had proposed that QPOs could represents a 
resonance in the coordinate frequencies  given by general relativistic motions close the black hole.


\section{Conclusion}
\label{conclusion}

The oscillations of fluid confined by gravity and centrifugal forces can lead to rich 
behavior under a variety of astrophysical scenarios, QPOs in X ray binaries. It is plausible that the 
characteristic frequencies of the collective motion  manifests itself in X-ray luminosity variation.
Under quite general considerations we have shown that simple, acoustic modes which have their origin 
in free particle oscillations are modified by the hydrodynamics and can couple to one another 
in non  linear fashion. The locking of excited modes could explain the fact that the 
observed QPO frequencies drift over considerable ranges while still arising from resonant 
interactions. We believe that their signature is present in the observations of accretion 
flows in strong gravitational fields, and will allow for the measurement of important 
parameters related to the black hole.
The strong- field effects of general relativity and in particular metric properties of
space-time around the black hole make the excitation of a 3:1 or 3:2 anharmonic epicyclic 
resonance, driven by orbital motion whose orbital frequency may be imprinted on the X-ray  
flux as a fairly prominent QPOs.

We have shown here that the tori and shock wave created on the accretion disk around the black hole 
produces oscillatory behavior. These types of structures constitute radial pressure gradients. So 
the pressure gradients themselves can permit the existence of discrete modes.
As a consequence of these dynamics, X-rays can be emitted by black hole 
binary system. The results of these different dynamical structures on the accretion disk 
responsible for QPOs are discussed in detail. The observed frequency ratios are consistent with the 
computed numerical values. On the other hand, our numerical results exhibits different commensurate 
frequencies in $4:1$, $5:1$..., seen in Figs.\ref{fig:NA:3:1} and \ref{fig:Num:3:1}.


\end{document}